\documentclass[aps,pre,showpacs,twocolumn,superscriptaddress]{revtex4-1} 	%looks like final product
\usepackage{graphicx, amsmath, amssymb,color}

\def \lvec{(\kern-.26em(}

\begin{document}

%opening
\title{The Thermodynamic Uncertainty Relation in Biochemical Oscillations}
\author{Robert Marsland III}
\affiliation{Department of Physics, Boston University, 590 Commonwealth Avenue, Boston, MA 02215}
\email[Email: ]{marsland@bu.edu}
\author{Wenping Cui}
\affiliation{Department of Physics, Boston University, 590 Commonwealth Avenue, Boston, MA 02215}
\affiliation{Department of Physics, Boston College, 140 Commonwealth Avenue, Chestnut Hill, MA 02467}
\author{Jordan M. Horowitz}
\affiliation{Physics of Living Systems Group, Department of Physics, Massachusetts Institute of Technology, 400 Technology Square, Cambridge, MA 02139}
\affiliation{Department of Biophysics, University of Michigan, Ann Arbor, MI, 48109}
\affiliation{Center for the Study of Complex Systems, University of Michigan, Ann Arbor, MI 48104}
\date{\today}
\begin{abstract}
Living systems regulate many aspects of their behavior through periodic oscillations of molecular concentrations, which function as ``biochemical clocks.'' The chemical reactions that drive these clocks are intrinsically stochastic at the molecular level, so that the duration of a full oscillation cycle is subject to random fluctuations. Their success in carrying out their biological function is thought to depend on the degree to which these fluctuations in the cycle period can be suppressed. Biochemical oscillators also require a constant supply of free energy in order to break detailed balance and maintain their cyclic dynamics. For a given free energy budget, the recently discovered `thermodynamic uncertainty relation' yields the magnitude of period fluctuations in the most precise conceivable free-running clock. In this paper, we show that computational models of real biochemical clocks severely underperform this optimum, with fluctuations several orders of magnitude larger than the theoretical minimum.  We argue that this suboptimal performance is due to the small number of internal states per molecule in these models, combined with the high level of thermodynamic force required to maintain the system in the oscillatory phase. We introduce a new model with a tunable number of internal states per molecule, and confirm that it approaches the optimal precision as this number increases. 
\end{abstract}

\maketitle

%To Do:
%-- NEW FRAME: Precision is way lower than maximum -- observed by Barato in AI, confirmed by us in KaiC. Why is this? Location of phase transition.
%-- Process toy data with new method
%-- Change legends to Delta S_cyc
%-- Add more annotation to cycle diagram
%-- Plot just one choice of parameters, but include many runs on the J vs. T plot
%-- Make histograms of several cuts through the runs
%-- Look up experiment papers on KaiC, to argue directly from the numbers that the bound is irrelevant

\begin{figure*}
	\includegraphics[width=17cm]{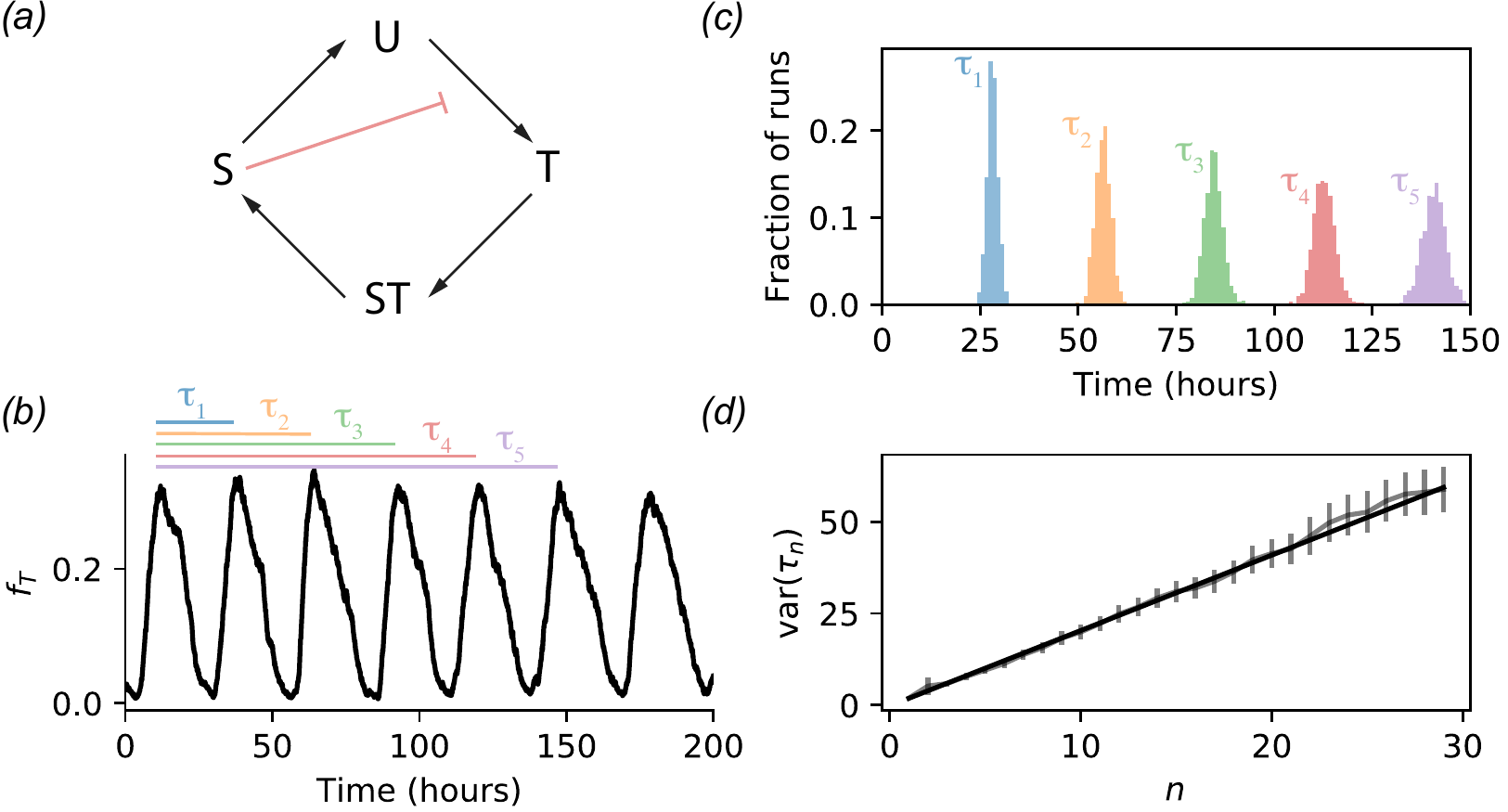}
	\caption{ {\bf Coherent cycles in a biochemical oscillator.} \emph{(a)} Schematic of the KaiC biochemical oscillator: This simplified diagram shows four different internal states of the KaiC molecule, labeled U, T, S and ST (unphosphorylated, phosphorylated on threonine, phosphorylated on serine, and phosphorylated on both residues). The molecules execute cycles around these four states in the indicated direction. They interact with each other via additional molecular components, in such a way that molecules in state S slow down the forward reaction rate for other molecules in state U. \emph{(b)} Simulated trajectory from a detailed kinetic model of the KaiC system (adapted from \cite{Paijmans2017}) with 360 interacting KaiC hexamers. The state space of this system is described by the list of copy numbers of all the molecular subspecies. Here we have projected the state onto a two-dimensional plane, spanned by the copy numbers of monomers in states T and S. The inset gives a magnified view of a small portion of the plot, showing the discrete reaction steps caused by single phosphorylation and dephosphorylation events. \emph{(c)} Time-evolution of the fraction $f_T$ of monomers in state T. \emph{(d)} Histograms of the time $\tau_n$ required for 1,200 independent sets of 360 interacting KaiC hexamers to complete $n$ collective cycles, for $n = 1$ through 10. \emph{(e)} Variances ${\rm var}(\tau_n)$ of the histograms as a function of $n$. Error bars are bootstrapped 95\% confidence intervals. Black line is a linear fit, with slope $D= 2.05\pm 0.05\ {\rm hours}^2$.}
	\label{fig:diagram}
\end{figure*}

Many living systems regulate their behavior using an internal ``clock,'' synchronized to the daily cycles of light and darkness. In the past 15 years, the isolation of the key components of several bacterial circadian clocks has opened the door to systematic and quantitative study of this phenomenon. In particular, a set of three proteins purified from the bacterium \emph{Synechococcus elongatus} are capable of executing sustained periodic oscillations \emph{in vitro} when supplied with ATP \cite{Nakajima2005}. One of the proteins,  KaiC, executes a cycle in a space of four possible phosphorylation states, as illustrated in figure \ref{fig:diagram}. This cycle is coupled to the periodic association and dissociation from the other two proteins, KaiA and KaiB. 

%This \emph{in vitro} system lacks a mechanism for phase-locking to periodic changes in light intensity, which is a crucial part of the biological function of the proteins \emph{in vivo} \cite{Pittayakanchit2018}. But it still allows for controlled study of the intrinsic dynamics of the free-running clock, which play an important role in determining the efficiency of phase-locking \cite{Fei2018}. In particular, one is interested in the size of fluctuations in the clock period, which can be measured by counting the number of periods $\mathcal{N}$ required for an ensemble of initially synchronized clocks to become randomly distributed around the clock cycle \cite{Cao2015,Barato2017,Fei2018}.

Steady oscillations break detailed balance, and must be powered by a chemical potential gradient or other free energy source. In this system and in related experiments and simulations, it is commonly observed that the oscillator precision decreases as this thermodynamic driving force is reduced \cite{Cao2015}. At the same time, recent theoretical work indicates that the precision of a generic biochemical clock is bounded from above by a number that also decreases with decreasing entropy production per cycle \cite{Barato2015,Barato2016,Gingrich2016,Pietzonka2016c,Barato2017,Gingrich2017,Horowitz2017,Wierenga2018}. This has led to speculation that this universal bound may provide valuable information about the design principles behind real biochemical clocks. 

So far, most discussion of this connection has focused on models with cyclic dynamics hard-wired into the dynamical rules \cite{Barato2017,Marsland2018,Wierenga2018}. But real biochemical oscillators operate in a high-dimensional state space of concentration profiles, and the cyclic behavior is an emergent, collective phenomenon \cite{Cao2015,Barato2017,Nguyen2018a}. These oscillators typically exhibit a nonequilibrium phase transition at a finite value of entropy production per cycle. As this threshold is approached from above, the oscillations become more noisy due to critical fluctuations~\cite{Nguyen2018a,Qian2000,Herpich2018,Lee2018}. Below the threshold, the system relaxes to a single fixed point in concentration space, with no coherent oscillations at all. In some systems, the precision may still be well below the theoretical bound as the system approaches this threshold. In these cases, the precision will never come close to the bound, for any size driving force. 

As we show in Section \ref{sec:underperform}, computational models of real chemical oscillations typically fall into this regime, never approaching to within an order of magnitude of the bound. Macroscopic \emph{in vitro} experiments on the KaiABC system perform even worse, remaining many orders of magnitude below the bound. Previous theoretical work suggests that the performance could be improved by increasing the number of reactions per cycle at fixed entropy production, and by making the reaction rates more uniform \cite{Barato2015,Pietzonka2016,Wierenga2018}. In Section \ref{sec:bound}, we elaborate on these ideas, introducing an effective number of states per cycle and showing how the relationship of this quantity to the location of the phase transition threshold controls the minimum distance to the bound. In Section \ref{sec:toy}, we introduce a new model based on these design principles, with nearly uniform transition rates in the steady state and with a tunable number of reactions per cycle. We show that this model approaches the optimal precision as the number of reaction steps per cycle grows. 

\section{Effective number of states and critical entropy production control distance to thermodynamic bound}
\label{sec:bound}

As illustrated in figure \ref{fig:diagram}\emph{(a)}, a KaiC monomer has two phosphorylation sites, one on a threonine residue (T) and one on a serine (S), giving rise to four possible phosphorylation states \cite{Nakajima2005}. The monomer also has two ATP-binding pockets, and forms hexamers that collectively transition between two conformational states \cite{phong2013robust,tseng2017structural}. All these features are important for the dynamics of the system, and have been incorporated into a thermodynamically consistent computational model that correctly reproduces the results of experiments performed with purified components \cite{Paijmans2017}. In particular, the ATP hydrolysis rate in one of the binding pockets has been shown to be essential for determining the period of the circadian rhythms \cite{Terauchi2007,phong2013robust,tseng2017structural}. Although our simulations will use the detailed model just mentioned, the simplified schematic in figure \ref{fig:diagram} highlights only the features of the model that we will explicitly discuss: the fact that each molecule can execute a directed cycle among several internal states (pictured as black arrows), and the fact that the state of one molecule affects transition rates of the others (suggestively represented as a red inhibition symbol).

Fluctuations in the time required for a simplified model of a single KaiC hexamer to traverse the reaction cycle have recently been studied in \cite{Barato2017}. But the biological function of this clock demands more than precise oscillations of isolated molecules; rather, it has evolved to generate oscillations in the \emph{concentrations} of various chemical species. The concentrations are global variables, which simultaneously affect processes throughout the entire cell volume. These global oscillations can still be described by a Markov process on a set of discrete states, but with a very different topology from the unicyclic network of an isolated monomer. For a well-mixed system, each state can be labeled by a list of copy numbers of all molecules in the reaction volume as shown in figure \ref{fig:diagram}\emph{(b)}, with each distinct internal state counted as a different kind of molecule. 

In the KaiC system, molecules in one of the phosphorylation states can suppress further phosphorylation of other molecules, by sequestering the enzyme (KaiA) required to catalyze the phosphorylation. This mechanism can stably synchronize the progress of all the molecules around the phosphorylation cycle, slowing down the ones that happen to run too far ahead of the rest. This is crucial for the maintenance of sustained oscillations in the concentration of free KaiA and of each of the four forms of KaiC. Figure \ref{fig:diagram}\emph{(c)} shows a sample trajectory of the concentration of one of the KaiC phosphorylation states in the detailed computational model mentioned above \cite{Paijmans2017}. 

Unlike the cycles of an idealized mechanical clock, the period $\tau_1$ of these oscillations is subject to random fluctuations, due to the stochastic nature of the underlying chemical reactions. The precision can be quantified by considering an ensemble of identical reaction volumes that are initially synchronized. The histogram of times $\tau_n$ for each molecule to complete $n$ cycles will widen as $n$ increases and the clocks lose their initial synchronization, as illustrated in figure \ref{fig:diagram}\emph{(d)}. When the width exceeds the mean period $T \equiv \langle \tau_1\rangle$, the clocks are totally desynchronized. This leads to a natural measure of the precision of the clock in terms of the number of coherent cycles $\mathcal{N}$ that take place before the synchronization is destroyed.

To measure $\mathcal{N}$ in a systematic way, we first note that the variance ${\rm var}(\tau_n) = Dn$, for some constant of proportionality $D$, as illustrated in figure \ref{fig:diagram}\emph{(e)}. This is exactly true in a renewal process~\cite{Ptaszynski2018}, such as the isolated KaiC monomer, where each period is an independent random variable (cf. \cite{Wierenga2018}). It remains asymptotically valid for arbitrarily complex models in the large $n$ limit, as long as the correlation time is finite. The number of cycles required for the width $\sqrt{{\rm var}(\tau_n)}$ of the distribution to reach the average period $T$ is therefore given by
\begin{align}
\mathcal{N} \equiv \frac{T^2}{D}.
\end{align}

Any chemical oscillator must be powered by a detailed-balance-breaking thermodynamic driving force that generates a positive average rate of entropy production $\dot{S}$. The number of coherent cycles is subject to a universal upper bound as a function of $\dot{S}$, holding for arbitrarily complex architectures \cite{Barato2015,Barato2016,Gingrich2016,Pietzonka2016c,Barato2017,Horowitz2017,Dechant2018}. The bound says that $\mathcal{N}$ is never greater than half the entropy production per cycle $\Delta S \equiv \dot{S}T$ (setting Boltzmann's constant $k_B=1$ from here on)~\cite{Gingrich2017}:
\begin{align}
\label{eq:bound}
\mathcal{N} \leq \frac{\Delta S}{2}.
\end{align}
The validity of this bound depends on the proper definition of $\mathcal{N}$, which in our formulation also depends on the definition of $\tau_n$. Determining $\tau_n$ is a subtle matter for systems of interacting molecules.  Our solution is presented in detail in the Appendix, but it always roughly corresponds to the peak-to-peak distance in figure \ref{fig:diagram}\emph{(c)}.

\begin{figure}
	\includegraphics[width=8cm]{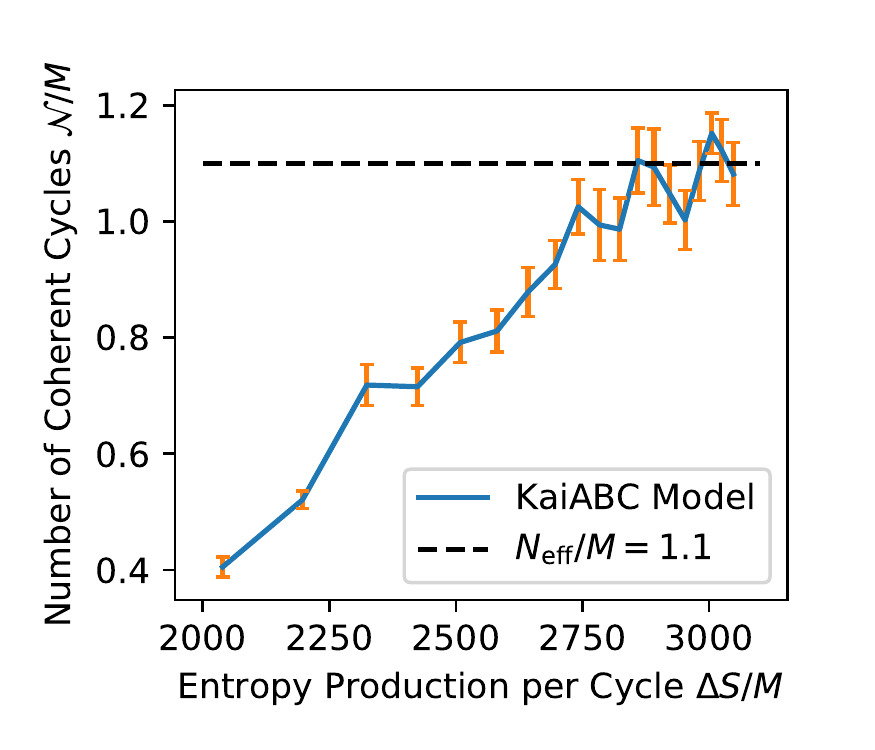}
	\caption{ {\bf Number of coherent oscillations saturates as $\Delta S\to\infty$.} The number of coherent cycles $\mathcal{N}$ is plotted as a function of the entropy production per cycle $\Delta S$ for the KaiC model discussed in figure \ref{fig:diagram} above. Both axes are scaled by the system size $M$, so that all quantities are molecular-scale values. Error bars are $\pm 1$ standard deviation, estimated with the bootstrap procedure described in the Appendix. The black dotted line is the estimated $\Delta S\to\infty$ limit $\mathcal{N} = N_{\rm eff}$. See Appendix and \cite{Paijmans2017} for model parameters.}
	\label{fig:Neff}
\end{figure}

As $\Delta S \to \infty$, Equation (\ref{eq:bound}) says that $\mathcal{N}$ is also allowed to become arbitrarily large. But as the entropy released in the reactions coupled to the driving force increases, detailed balance implies that the reverse reaction rates tend towards zero. Once the reverse rates are negligible compared to the other time scales of the problem, these reactions can be ignored, and further changes in $\Delta S$ produce no effect. In any given biochemical model, therefore, $\mathcal{N}$ approaches some finite value as $\Delta S \to \infty$ (as was already noted in \cite{Cao2015}), which depends on the network topology and the rest of the reaction rates.  
We can see this in our detailed computational model in figure~\ref{fig:Neff}.
For unicyclic networks in particular, where the  topology is a single closed cycle like the isolated KaiC monomer, the maximum possible value for this asymptote is the number of states $N$ \cite{David1987,Marsland2018}. By analogy, we will refer to the $\Delta S\to\infty$ limit of $\mathcal{N}$ for any model as the effective number of states per cycle $N_{\rm eff}\equiv\lim_{\Delta S\to\infty}{\mathcal N}$. For an oscillator built from coupled cycles of internal states, such as the KaiC system, $N_{\rm eff}$ reaches its maximum value when the dynamics constrain the oscillations to a single path through concentration space, and when all reaction rates along this path are equal. In this case, the dynamics are equivalent to a single ring of $NM$ states, where $N$ is the number of internal states per molecule and $M$ is the number of molecules. This upper bound on $N_{\rm eff}$ can be easily computed for any model or experiment from a basic knowledge of the component parts. 

In all five models we will analyze below, $\mathcal{N}$ monotonically increases as a function of $\Delta S$. The existence of the finite $\Delta S \to\infty$ limit thus implies that $\mathcal{N}$ can only approach the thermodynamic bound of Equation (\ref{eq:bound}) when $\Delta S < \Delta S_{\rm b} \equiv 2 N_{\rm eff}$. But the collective oscillations of these models also exhibit a nonequilibrium phase transition as a function of $\Delta S$, whose critical behavior has  recently been studied \cite{Cao2015,Nguyen2018}. In the thermodynamic limit, the inverse precision $1/\mathcal{N}$ diverges as $\Delta S$ approaches a critical value $\Delta S_{\rm c}$ from above, in a way that depends on the architecture of the reaction network. Below $\Delta S_{\rm c}$, there are no collective oscillations, and the concentrations relax to a single fixed point. Since the oscillations cease to exist below $\Delta S_{\rm c}$, the bound is only relevant for $\Delta S > \Delta S_{\rm c}$. 
Combining these two observations, we see that models with $\Delta S_{\rm b} < \Delta S_{\rm c}$ can never approach the thermodynamic bound. 

\section{Models of real chemical oscillators severely underperform the bound}
\label{sec:underperform}

\begin{figure}
	\includegraphics[width=8cm]{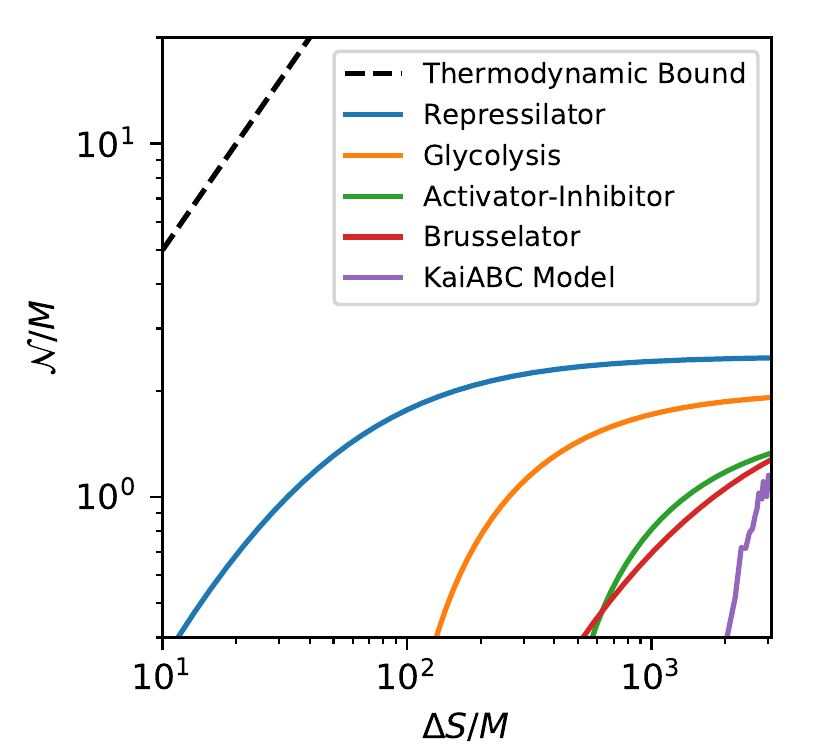}
	\caption{ {\bf Models of collective oscillations compared with thermodynamic bound.} Same as figure \ref{fig:Neff}, but including all four models studied in \cite{Cao2015}. The black dotted line is the thermodynamic bound $\mathcal{N} = \Delta S/2$. Curves for the first four models are phenomenological fits obtained in \cite{Cao2015}.}
	\label{fig:models}
\end{figure}

Cao \emph{et.\ al.}~recently measured $\mathcal{N}$ as a function of $\Delta S$ in computational models of four representative chemical clock architectures: activator-inhibitor, repressilator, Brusselator, and the glycolysis network \cite{Cao2015}. The data for all four models produced an acceptable fit to a four-parameter phenomenological equation, which is reproduced in the Appendix along with the parameter values obtained by Cao \emph{et.\ al.}~for each model. In figure \ref{fig:models}, we plot these phenomenological curves and the thermodynamic bound of Equation (\ref{eq:bound}). We also obtained $\mathcal{N}$ and $\Delta S$ for a detailed model of the KaiC system based on \cite{Paijmans2017} as described in the Appendix, with the parameters obtained in that paper by extensive comparison with experimental data, for twenty values of the ATP/ADP ratio.

The values of $N_{\rm eff}$, $\Delta S_{\rm b}$ and $\Delta S_{\rm c}$ can be estimated directly from figure \ref{fig:models}, by noting where each curve saturates and where it drops to zero. Both axes are scaled by the system size $M$, which equals the number of KaiC hexamers for the KaiC model, and the number of kinases in the activator-inhibitor model. The other three models lack a direct physical interpretation of $M$, since there are no conserved molecular species, but it still defines a generic molecular scale. For any physically reasonable model, $\mathcal{N}$ is expected to be an extensive parameter, proportional to $M$, as is $\Delta S$. This has been confirmed numerically for a number of models, and appears to break down significantly only in the immediate vicinity of the critical point \cite{Cao2015,Nguyen2018,Lee2018}. The models plotted here have $N_{\rm eff}/M \approx 2$, which is reasonable for molecules that only have a few internal states and highly non-uniform reaction rates. 

But $N_{\rm eff}/M \approx 2$ implies that $\Delta S_{\rm b}/M \approx 4$, which means that the entropy production per molecule per cycle must be less than 4 for the thermodynamic bound to become relevant. This is a very small number even by biochemical standards, equal to the entropy change from forming four hydrogen bonds between protein residues in solution. The activator-inhibitor, Brusselator, and glycolysis models have phase transitions at $\Delta S_{\rm c}/M$ values of 360, 100.4 and 80.5, respectively, under the parameter choices of \cite{Cao2015}. They all exceed $\Delta S_{\rm b}/M$ by at least an order of magnitude, guaranteeing that the precision can never come close to the thermodynamic bound. The KaiC model appears to have $\Delta S_{\rm c}/M \sim 1,000$ and $N_{\rm eff}/M = 1.1$, so that $\Delta S_{\rm c}$ exceeds $\Delta S_{\rm b}$ by more than two orders of magnitude. The only model with $\Delta S_{\rm b} > \Delta S_{\rm c}$ is the repressilator model, where $\Delta S_{\rm c}/M = 1.75$ and $\Delta S_{\rm b} \approx 4$. But even here, the critical fluctuations begin to severely degrade the precision when $\Delta S$ is still much greater than $\Delta S_{\rm b}$. 

Estimates of $\mathcal{N}$ can also be extracted directly from experiments, as shown by Cao \emph{et.\ al.} for an \emph{in vitro} reconstitution of the KaiC system with purified components in a macroscopic reaction volume \cite{Cao2015}. They analyzed timeseries data from a set of experiments at different ATP/ADP ratios, and fit their phenomenological equation to describe $\mathcal{N}$ as a function of this ratio. As we show in the Appendix, this fit implies that $\Delta S_{\rm c}/M\sim 1,000$, consistent with our model results. The $\Delta S\to\infty$ asymptote of the fit, however, gives $N_{\rm eff}/M \sim 10^{-11}$, which is astronomically small compared to the model prediction $N_{\rm eff}/M \approx 1$. This surprising result reflects the fact that the dominant sources of uncertainty in these macroscopic experiments are fluctuations in temperature and other environmental perturbations, rather than the intrinsic stochasticity of the reaction kinetics. Since these fluctuations are independent of the system size, their effect is inflated when we divide by the number of hexamers $M \sim 10^{14}$ in a 100 $\mu$L reaction volume at 1 $\mu$M concentration. The only way to observe the effect of intrinsic stochasticity in such a noisy environment is to decrease the reaction volume. Assuming that the minimum contribution of the external noise to $\mathcal{N}$ remains fixed at the value of 500 found in the experiments, and that the intrinsic contribution is of order $N_{\rm eff} \approx M$ as given by the model, we find that $M = 500$ hexamers is the system size at which the intrinsic fluctuations become detectable. At 1 $\mu$M concentration of hexamers, the corresponding reaction volume is about $1 \mu$m$^3$, the size of a typical bacterial cell.

Because of this separation of scales, the apparent divergence of $1/\mathcal{N}$ at a critical value of the ATP/ADP ratio in the experiments is probably due to the expected divergence of susceptibility at the critical point, which makes the oscillation period increasingly sensitive to environmental fluctuations as the ATP/ADP ratio is reduced. In any case, this analysis suggests that an important design consideration for oscillators in living systems is robustness against external perturbations, as recently explored in \cite{Monti2018,pittayakanchit2018biophysical,delJunco2018}.

\section{Toy model with variable number of states can saturate the bound}
\label{sec:toy}

\begin{figure}
	\includegraphics[width=8cm]{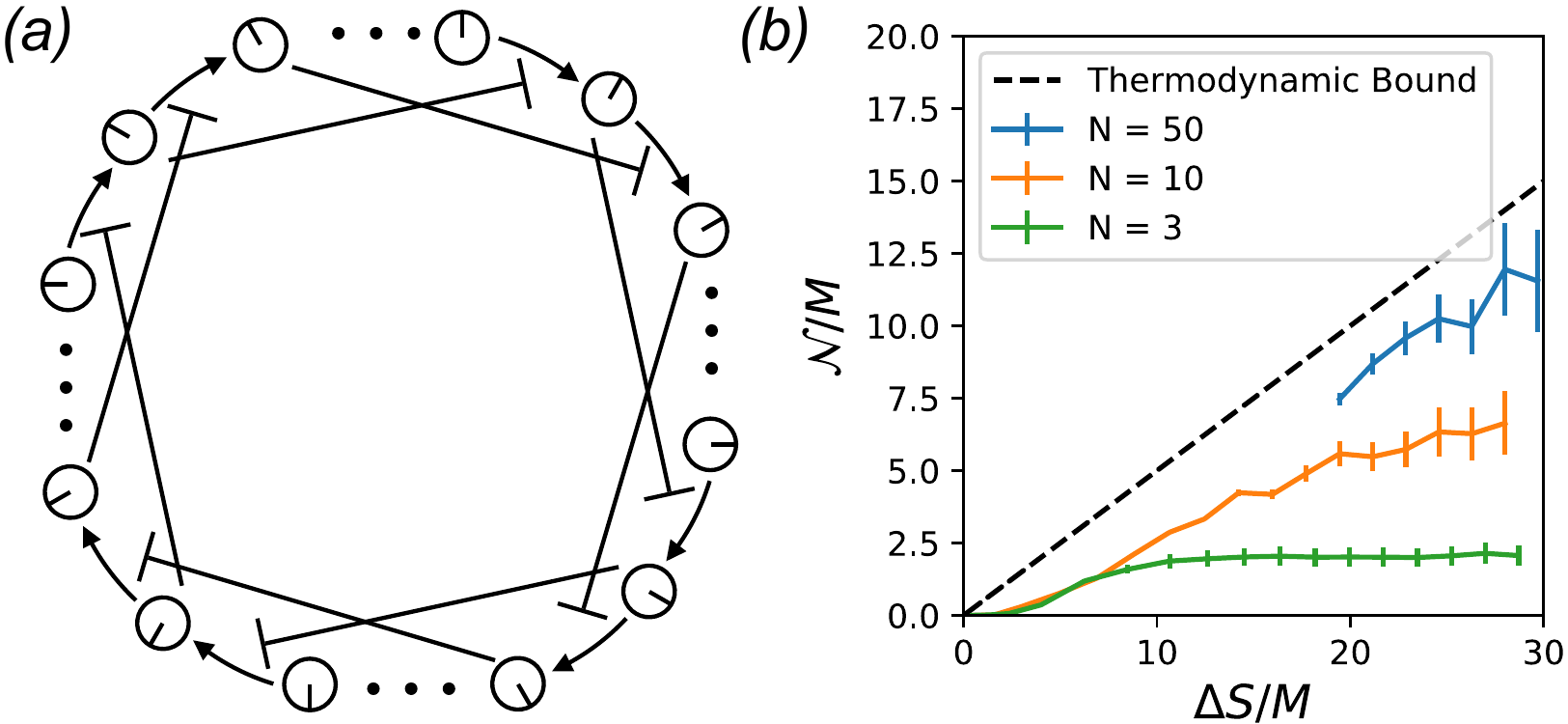}
	\caption{ {\bf Symmetric toy model compared with thermodynamic bound.} (a) Schematic of toy model inspired by the KaiC system. Each protein has $N$ distinct conformations whose transitions are arranged in a ring topology with a net clockwise drift. Each protein suppresses the transition rate for proteins further along in the circulation around the ring. (b) $\mathcal{N}/M$ versus $\Delta S/M$ in this model for three different values of the number of internal states $N$. Error bars are $\pm 1$ standard deviation, estimated with the same bootstrap procedure used in figure \ref{fig:Neff}. See Appendix for model details and parameters.}
	\label{fig:toy}
\end{figure}

The failure of all five of these models to approach the thermodynamic bound raises the question of whether it is possible in principle for any chemical oscillator to do so.
Put another way, whether it is possible to simultaneously achieve a large enough $N_{\rm eff}/M$ and small enough $\Delta S_{\rm c}/M$. In a simple unicyclic network of $N$ states, $N_{\rm eff} = N$ when all the reaction rates in the forward direction are equal, and so one can always approach arbitrarily close to the bound by increasing $N$. But a chemical oscillator cannot have uniform rates, since the transition rates of each molecule have to change based on the states of the others in order to achieve collective oscillations. Furthermore, it is not known how changing the number of internal states affects $\Delta S_{\rm c}/M$, and so it is not obvious whether $\Delta S_{\rm c} < \Delta S_{\rm b}$ is achievable at all. 

To answer this question, we devised a new model loosely inspired by the KaiC model of figure \ref{fig:diagram}, with $M$ interacting molecules each containing $N$ distinct internal states. Molecules in any one of these states suppress the transition rates for other molecules that are further ahead in the cycle, as illustrated in figure \ref{fig:toy} and described in detail in the Appendix. All internal states have the same energy, and each reaction carries the same fraction of the total thermodynamic force. 

In this highly symmetrized model, $\Delta S_{\rm c}/M$ can easily be reduced to between 2 and 3 by choosing a sufficiently high coupling strength, as shown in the Appendix. At the same time $N_{\rm eff}/M$ scales linearly with $N$, and can be made arbitrarily large by increasing this parameter. In figure \ref{fig:toy}, we plot $\mathcal{N}/M$ versus $\Delta S/M$ for three different values of $N$, and show that the data does indeed approach the thermodynamic bound as $N$ increases. This extends the validity of design principles obtained for unicyclic networks in various context to these collective dynamics: the rates should be made as uniform as possible, while the number of internal states is made as large as possible at fixed thermodynamic driving force \cite{Barato2017,Wierenga2018,Lang2014}. 

\section{Discussion}
The thermodynamic uncertainty relation is a powerful result with impressive universality.  It has been widely assumed that the relation should have some relevance for the evolution of biochemical oscillators. Based on data from experiments and extensive simulations with realistic parameters, we have argued that these oscillators typically underperform the bound by at least an order of magnitude. From a thermodynamic perspective, they are free to evolve higher precision without increasing their dissipation rate. 

We have also derived a simple criterion for estimating how closely a given oscillator can approach the thermodynamic bound, in terms of an effective number of states $N_{\rm eff}$ and the entropy production per cycle $\Delta S_{\rm c}$ at the onset of oscillatory behavior. For an oscillator composed of $M$ identical molecules with $N$ internal states, globally coupled through mass-action kinetics, we noted that $N_{\rm eff}\leq NM$, with equality only when all the cycles are perfectly synchronized, and when all reactions that actually occur have identical rates. Assuming that the number of coherent periods $\mathcal{N}$ is monotonic in the entropy production per cycle $\Delta S$, the thermodynamic bound can only be approached when $NM \geq N_{\rm eff} \gg \Delta S_{\rm c}$. 

To show that this criterion can in principle be satisfied by emergent oscillations of molecular concentrations, we devised a toy model that oscillates with less than 3 $k_B T$ of free energy per molecule per cycle and can contain an arbitrary number of internal states per molecule. But it is hard to imagine a biochemically plausible mechanism for sustained oscillations powered by the free energy equivalent of three hydrogen bonds per molecule per cycle. Certainly this could not be a phosphorylation cycle, since cleaving a phosphate group from ATP and releasing it into the cytosol dissipates about 20 $k_B T$ under physiological conditions \cite{moran2010snapshot}. Furthermore, we noted that the five models we analyzed all have an effective number of internal states per molecule $N_{\rm eff}/M$ of around 2. This number may be constrained by a trade-off with the complexity of the oscillator. For KaiC, the inter-molecular coupling required for collective oscillations is mediated by sequestration of KaiA at a particular point in the cycle. Implementing the symmetric $N$-state cycle of figure 4(a) in this way would require each transition to be catalyzed by a different molecule, and for each of these molecules to be selectively sequestered by the state a quarter-cycle behind the transition. Finally, biochemical oscillators are subject to other performance demands that may be more important than the number of coherent cycles of the free-running system. In particular, an oscillator's ability to match its phase to an external signal (e.g., the day/night cycle of illumination intensity) is often essential to its biological function, placing an independent set of constraints on the system's architecture \cite{pittayakanchit2018biophysical,chew2018high}. Entrainment efficiency has recently been shown to increase with dissipation rate in some models even when the free-running precision has saturated, providing an impetus for increasing the entropy production per cycle beyond what is required to achieve $\mathcal{N} = N_{\rm eff}$ \cite{fei2018design}.

\begin{acknowledgments}
We thank J. Paijmans for his help with adjustments to the KaiC simulation, and Y. Cao for valuable discussions on the technical details of reference \cite{Cao2015}. RM acknowledges Government support through NIH NIGMS grant 1R35GM119461.  JMH is supported by the Gordon and Betty Moore Foundation as a Physics of Living Systems Fellow through Grant No. GBMF4513. The computational work reported on in this paper was performed on the Shared Computing Cluster which is administered by Boston University’s Research Computing Services.
\end{acknowledgments}

\bibliographystyle{abbrv}
%\bibliography{library}

\section*{Appendix}
\subsection{Measuring the stochastic period}
The definition of the number of coherent cycles $\mathcal{N}$ depends on a prior notion of the $n$-cycle completion time $\tau_n$. In a unicyclic transition network, this time can be straightforwardly defined in terms of the integrated current $J$ through an arbitrarily chosen transition in the network. Each time the transition is executed in the forward direction, $J$ increases by 1, and each time it is executed in the reverse direction, $J$ decreases by 1. The $n$-cycle completion time $\tau_n$ is then naturally defined as the time when the system first reaches $J = n$, given that it was initialized in the state immediately adjacent to the measured transition in the positive direction \cite{Gingrich2017,Wierenga2018,Ptaszynski2018}.

\begin{figure}
	\includegraphics[width=8cm]{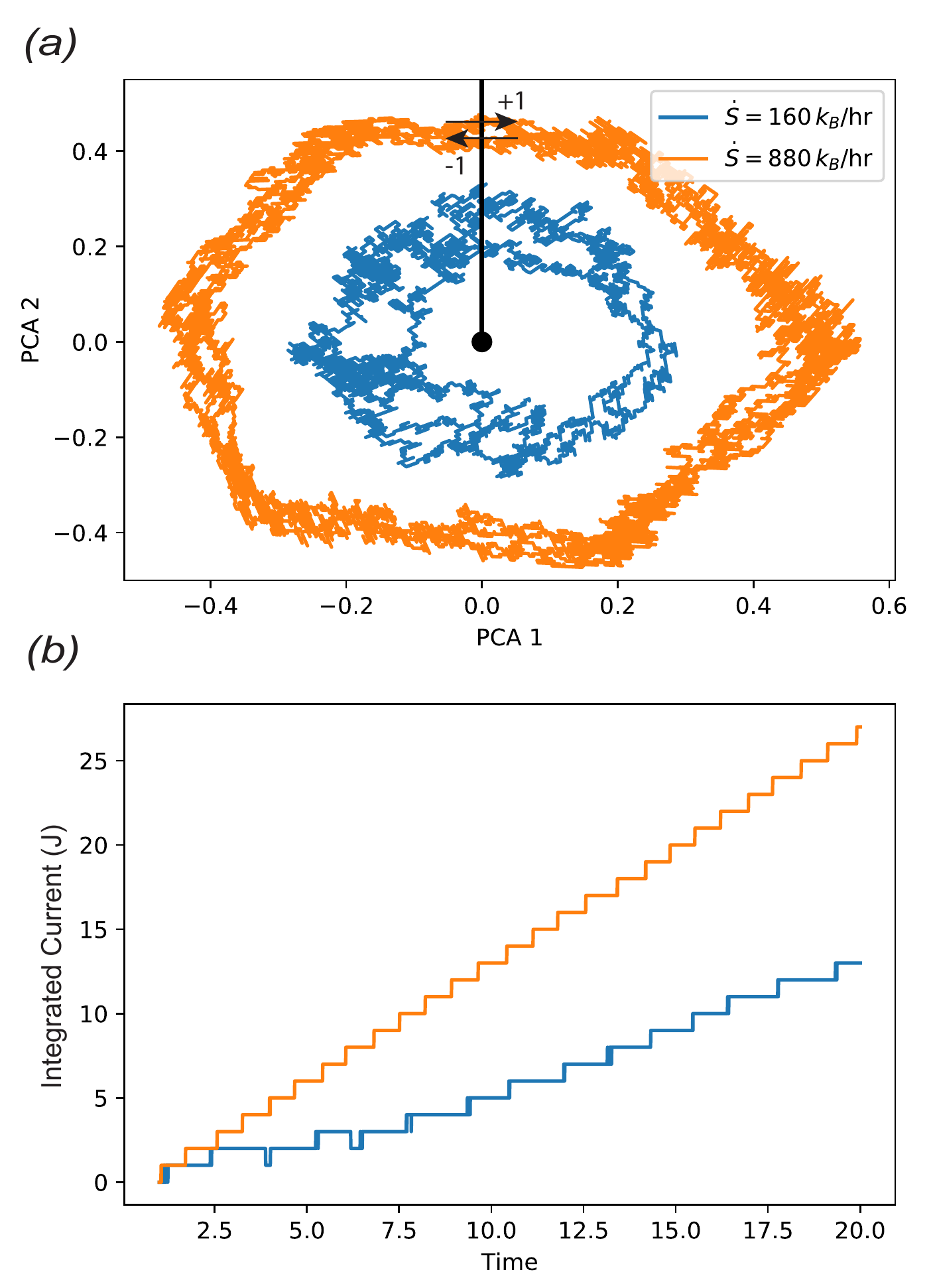}
	\caption{ {\bf Measuring the $n$-cycle completion time $\tau_n$} (a) Trajectories of the toy model of figure \ref{fig:toy} with $N = 6$ for two different values of the thermodynamic driving force, projected onto their first two principal components. A cut from the origin along the positive vertical axis provides the criterion for cycle completion. (b) Integrated current (net number of completed cycles) $J$ as a function of time for the same two trajectories. The first-passage time for a net increase $n$ in the integrated current defines the $n$-cycle completion time $\tau_n$.}
	\label{fig:winding}
\end{figure}

For a chemical oscillator, the definition is not so clear. One common approach is to fit the autocorrelation function of some observable to a sine wave with exponentially decaying amplitude. If the autocorrelation function exactly fits this functional form, then $\mathcal{N}$ can be obtained from the ratio of the decay time to oscillation period via a numerical conversion factor \cite{Cao2015}. One can also evaluate the ratio of imaginary to real parts of the leading eigenvalue of the rate matrix for the Master equation of the dynamics, which gives the same result when all the other eigenvalues are much smaller in amplitude \cite{Qian2000,Barato2017,Nguyen2018a}. 
While this ratio is conjectured to be bounded by the thermodynamic driving force powering the oscillations, it is not the approach we study here \cite{Barato2017}.
Instead, we note that the value of $\mathcal{N}$ generated by these preceding procedures only satisfies the hypotheses of the thermodynamic uncertainty relation under the specific conditions of a perfectly sinusoidal autocorrelation function (cf. \cite{Barato2017}).  For our analysis of the KaiC model and our new toy model, we instead utilize a definition of $\tau_n$ that treats the oscillations in the full concentration space as one large cycle.

Figure \ref{fig:winding} illustrates our procedure. We started by projecting the state of the system from the high-dimensional concentration space to two dimensions. We projected onto the plane that captured the largest percentage of the total variation in system state over a cycle, using a Principal Component Analysis (PCA) of a trajectory containing at least one full cycle (using the Python package scikit-learn \cite{scikit-learn}). In this plane, the oscillating trajectories describe a noisy ring, as shown in Figure \ref{fig:winding}a. Because the ring has a finite width, we cannot select a single transition to count the integrated current $J$. Instead, we draw a half-line starting from the middle of the ring, representing a half-hyperplane in the full state space, and include all the transitions cut by this hyperplane in the current. Each time the line is crossed in the clockwise direction, $J$ increases by 1, and each time it is crossed in the counterclockwise direction, $J$ decreases by 1. Sample traces of $J(t)$ are plotted in figure \ref{fig:winding} \emph{(b)}. The $n$-cycle completion time $\tau_n$ can now be defined as before, measuring the first-passage time for reaching $J = n$. These definitions fulfill the hypotheses of the thermodynamic uncertainty relations for currents and for first passage times \cite{Barato2015,Gingrich2016,Gingrich2017}. In the notation of \cite{Gingrich2016,Gingrich2017}, they correspond to setting $d(y,z)=1$ for all transitions from $y$ to $z$ cut by the hyperplane, and $d(y,z) = 0$ for all other transitions. Note that the uncertainty relations are obtained in the $J \to\infty$ limit, where initial conditions are irrelevant, but for our numerical analysis we chose special initial conditions that gave rapid convergence to the asymptotic form. Specifically, we employed a conditional steady-state distribution over states adjacent to the hyperplane. This was achieved by running the simulation longer than the relaxation time, and then starting the counter at $J=0$ the next time the hyperplane was crossed. 

\subsection{Phenomenological fits from reference \cite{Cao2015}}
The extensive numerical simulations performed by Cao \emph{et.\ al.} on four different models of chemical oscillators can be summarized by the parameters of a phenomenological fitting function
\begin{align}
\label{eq:fitting}
\frac{\mathcal{N}}{M} = \left[ A + B\left(\frac{\Delta S-\Delta S_{\rm c}}{M}\right)^\alpha \right]^{-1}
\end{align}
with four parameters $A, B, \Delta S_{\rm c}$ and $\alpha$. The exponent $\alpha$ is always negative, so $\mathcal{N}/M$ goes to zero as $\Delta S \to \Delta S_{\rm c}$. (To convert from their notation to ours, use $V\to M$, $W_0\to B$, $C\to A$, $W_c \to \Delta S_{\rm c}/M$, $D/T \to \mathcal{N}^{-1}$). The parameters for these fits are given in the figure captions of \cite{Cao2015}, and are reproduced in the following table:
\begin{center}
\begin{tabular}{l|c|c|c|c}
Model	& $A$ & $B$ & $\Delta S_{\rm c}/M$ & $\alpha$\\
\hline
Activator-Inhibitor & 0.6 & 380 & 360 & -0.99\\
Repressilator & 0.4 & 25.9 & 1.75& -1.1\\
Brusselator & 0.5 & 846 & 100.4 & -1.0\\
Glycolysis & 0.5 & 151.4 & 80.5 & -1.1
\end{tabular}
\end{center}
Note that $A$ controls the $\Delta S\to\infty$ asymptote, and is equal to $N_{\rm eff}^{-1}$. 

\subsection{Analysis of experimental data}
Cao \emph{et.\ al.} also analyze experimental data on the KaiC system, extracting the ratio of decay time to oscillation period from fits to the autocorrelation function at different values of the ATP/ADP ratio. They fit Equation (\ref{eq:fitting}) above to the data, but using $\ln ({\rm [ATP]}/{\rm [ADP]})$ instead of $\Delta S/M$, and without converting the autocorrelation ratio to $\mathcal{N}$ or dividing by volume. 

To compare these results to the thermodynamic bound, we first had to convert from the logarithm of the ATP/ADP ratio to entropy production per cycle. In the text, they estimate that the critical value $\ln ({\rm [ATP]}/{\rm [ADP]})_{\rm c} = -1.4$ obtained from the fit corresponds to an entropy production per ATP hydrolysis event of 10.6. Combining this with the measurement from \cite{Terauchi2007} of 16 hydrolysis events per cycle per KaiC monomer, we find a critical entropy production per cycle per hexamer of $\Delta S_{\rm c}/M \approx 10.6\times 16 \times 6 \approx 1,020$.

Next, we had to convert the observed ratio of decay time/period to $\mathcal{N}/M$. Since the autocorrelation function was well fit by an exponentially decaying sinusoid, we applied the corresponding conversion factor $\mathcal{N} = 2\pi^2 \frac{\tau}{T}$ where $T$ is the period and $\tau$ is the decay time \cite[Eq. 2]{Cao2015}. We then estimated the number of hexamers $M \approx 3\times 10^{13}$ using the KaiC monomer concentration of 3.4 $\mu$M reported with the original publication of the data \cite{Rust2011,Phong2013}, and the typical reaction volume in a 96-well plate of 100 $\mu$L. With these two conversions, we found that the $\Delta S\to\infty$ value of $\mathcal{N}/M$ was $2 \times 10^{-11}$. 

\subsection{Thermodynamically consistent KaiC model}
Paijmans \emph{et.\ al.} have recently developed a mechanistically explicit computational model of the KaiC oscillator \cite{Paijmans2017}. This model is particularly interesting from the theoretical point of view because it captures the extremely large dimensionality characteristic of real biochemical systems. Each KaiC hexamer contains six KaiC proteins, which each contain two nucleotide binding sites and two possible phosphorylation sites (``S'' and ``T'' from figure \ref{fig:diagram}). Each of these sites can be in one of two possible states (ATP-bound/ADP-bound, or phosphorylated/unphosphorylated). Furthermore, the whole hexamer can be in an ``active'' or an ``inactive'' conformation. Thus each hexamer has $(2\cdot 2\cdot 2 \cdot 2)^6 \cdot 2 = 2^{25}$ possible internal states. As noted in the main text, the state space for a well-mixed chemical system is the vector of concentrations of all molecular types. For the Paijmans \emph{et.\ al.} KaiC model, this vector therefore lives in a space of dimension $2^{25} = 3.4\times 10^7$. 

The original implementation of this model in \cite{Paijmans2017} lacked the reverse hydrolysis reaction ADP + P $\to$ ATP, which never spontaneously happens in practice under physiological conditions. To obtain full thermodynamic consistency, we added this reaction to the model with the assistance of one of the original authors. This required introducing a new parameter $\Delta G_0$, the free energy change of the hydrolysis reaction at standard concentrations. For all the simulations analyzed here, we chose $\Delta G_0$ and the concentration of inorganic phosphate [Pi] such that
\begin{align}
\frac{\rm [Pi]_0}{\rm [Pi]}e^{-\Delta G_0} = 10^8.
\end{align}
In other words, the entropy generated during a single hydrolysis reaction when nucleotide concentrations are equal ([ATP] = [ADP]) is $\Delta S_{\rm hyd} = \ln 10^8 \approx 18.4$.

Since the steady-state supply of free energy in this model comes entirely from the fixed nonequilibrium concentrations of ATP and ADP, we can measure the average rate of entropy production $\dot{S}$ by simply counting how many ATP molecules are hydrolyzed over the course of a long simulation, multiplying by $\Delta S_{\rm hyd}$, and dividing by the total time elapsed in the simulation.

All parameters other than $\Delta G_0$ are described and tabulated in the original publication \cite{Paijmans2017}, and the only parameter altered during our simulations was the ATP/ADP ratio.  

The revised C code and scripts for generating data can be found on GitHub at \url{https://github.com/robertvsiii/kaic-analysis}.

\subsection{Symmetric toy model}
We also developed our own abstract toy model to isolate the operating principles of the KaiC oscillator, and to check whether the thermodynamic bound could be saturated by a collective oscillator with the right design.

\begin{figure}
	\includegraphics[width=8cm]{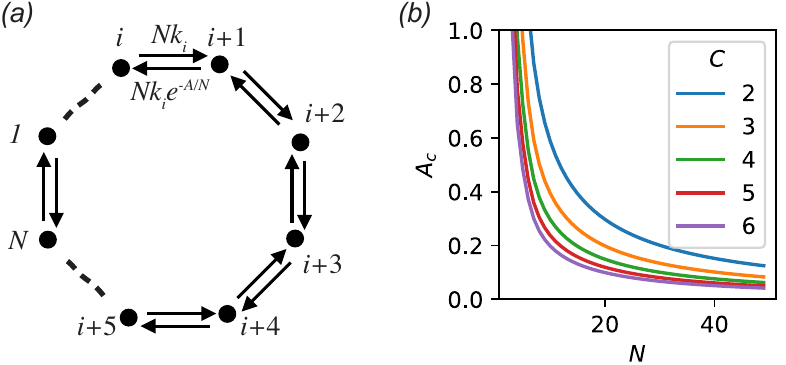}
	\caption{ {\bf Toy model with variable number of internal states} \emph{(a)} Transition rates for a single molecule with $N$ internal states. Multiple copies of the molecule are coupled together kinetically, by making $k_i$ depend on the fraction of molecules $f_i$ in each state. \emph{(b)} Dependence of critical affinity $A_c$ on $N$ for different values of the coupling $C$.}
	\label{fig:Ac}
\end{figure}

Consider a molecule with $N$ states, as sketched in figure \ref{fig:Ac} \emph{(a)}. Transitions are allowed from each state to two other states, such that the network of transitions has the topology of a ring. The rates for ``clockwise'' and ``counterclockwise'' transitions around this ring are $k^+ = Nk$ and $k^- = Nke^{-A/N}$, respectively, where $A$ is the cycle affinity. Under these definitions, the total entropy produced when one ring executes a full cycle is always equal to $A$. 

Now consider $M$ copies of this molecule in the same well-mixed solution with $M_i$ copies at position $i$, where $i$ increases in the ``clockwise'' direction from 1 to $N$. We can couple their dynamics together by allowing the bare rate $k$ to vary around the ring.
Specifically, we make the bare rate $k_i$ for transitions between states $i$ and $i+1$ depend on the occupancy fractions $f_j=M_j/M$ of all $N$ states: 
\begin{align}
k_i = \exp\left(-C\sum_j f_j \sin[2\pi(i-j)/N]\right)
\end{align}
The constant $C$ controls the strength of the coupling, and the rate for a uniform distribution over states is 1. This dependence of the rates on the $f_i$ mimics the effect of KaiA sequestration in the KaiC system. Recall that high occupancy of the inactive conformation of KaiC causes KaiA to be sequestered, slowing down nucleotide exchange in other hexamers, as illustrated in figure \ref{fig:diagram}. In this toy model, high occupancy of any one state slows down transition rates ahead of that state in the cycle, by up to a factor of $e^{-C}$ for transitions a quarter-cycle ahead. Due to the symmetry of our model, high occupancy of a given state also speeds up transition rates behind that state.

The data for figure \ref{fig:toy} was obtained with $C = 5$, for 18 values of $A$ from 1 to 30. Note that $\Delta S/M \approx A$, since all $M$ molecules execute approximately one cycle during a given oscillation period. 

We simulated this model using a Gillespie algorithm with the reaction rates specified above. The Python code can be found in the GitHub repository \url{https://github.com/robertvsiii/kaic-analysis}.

In the limit of infinite system size, the dynamics become deterministic, and are described by the following set of $N$ ODE's:
\begin{align}
\frac{df_i}{dt} = f_{i-1}k^+_{i-1} + f_{i+1}k^-_i - f_i(k^+_i + k^-_{i-1}),
\end{align}
with $k^+_i=Nk_i$ and $k^-_i=Nk_ie^{-A/N}$.
These equations always have a fixed point at the uniform state where $f_i = \frac{1}{N}$ for all $i$. The linearized dynamics around the uniform state can be written as:

\begin{align}
\frac{d\delta f_i}{dt} &= \frac{1}{N}\sum_j \left[ \frac{\partial k^+_{i-1}}{\partial f_j} +  \frac{\partial k^-_i}{\partial f_j}  -\left(\frac{\partial k^+_i}{\partial f_j} +\frac{\partial k^-_{i-1}}{\partial f_j} \right)\right]\delta f_j \nonumber\\
&+ (\delta f_{i-1} - \delta f_i)N + (\delta f_{i+1} - \delta f_i)N e^{-A/N}\\
&= \sum_j K_{ij} \delta f_j
\end{align}
where the last line defines the matrix $K_{ij}$. Oscillating solutions are possible when $K_{ij}$ acquires an eigenvalue with a positive real part, making this fixed point unstable. In figure \ref{fig:Ac} \emph{(b)}, we plot the critical affinity $A_c$ where these positive real parts first appear, as a function of the number of internal states $N$. We confirmed that the dominant pair of eigenvalues contains nonzero imaginary parts at $A = A_c$ for all points plotted, so that the transition is a true Hopf bifurcation to an oscillatory phase. 

Note that in the limit $A \to\infty, N\to\infty$, this model becomes identical to the irreversible limit of a fully-connected driven XY model.

\section{Simulations and Analysis}
To measure $\mathcal{N}$ and $\Delta S$ in the KaiC model and our new toy model, we generated an ensemble of trajectories for each set of parameters. Each ensemble of the KaiC model contained 1,200 trajectories, while each toy model ensemble contained 1,120 trajectories. Before collecting data, we initialized each trajectory by running the dynamics for longer than the empirically determined relaxation time of the system, in order to obtain a steady-state ensemble. 

\begin{figure*}
	\includegraphics[width=12cm]{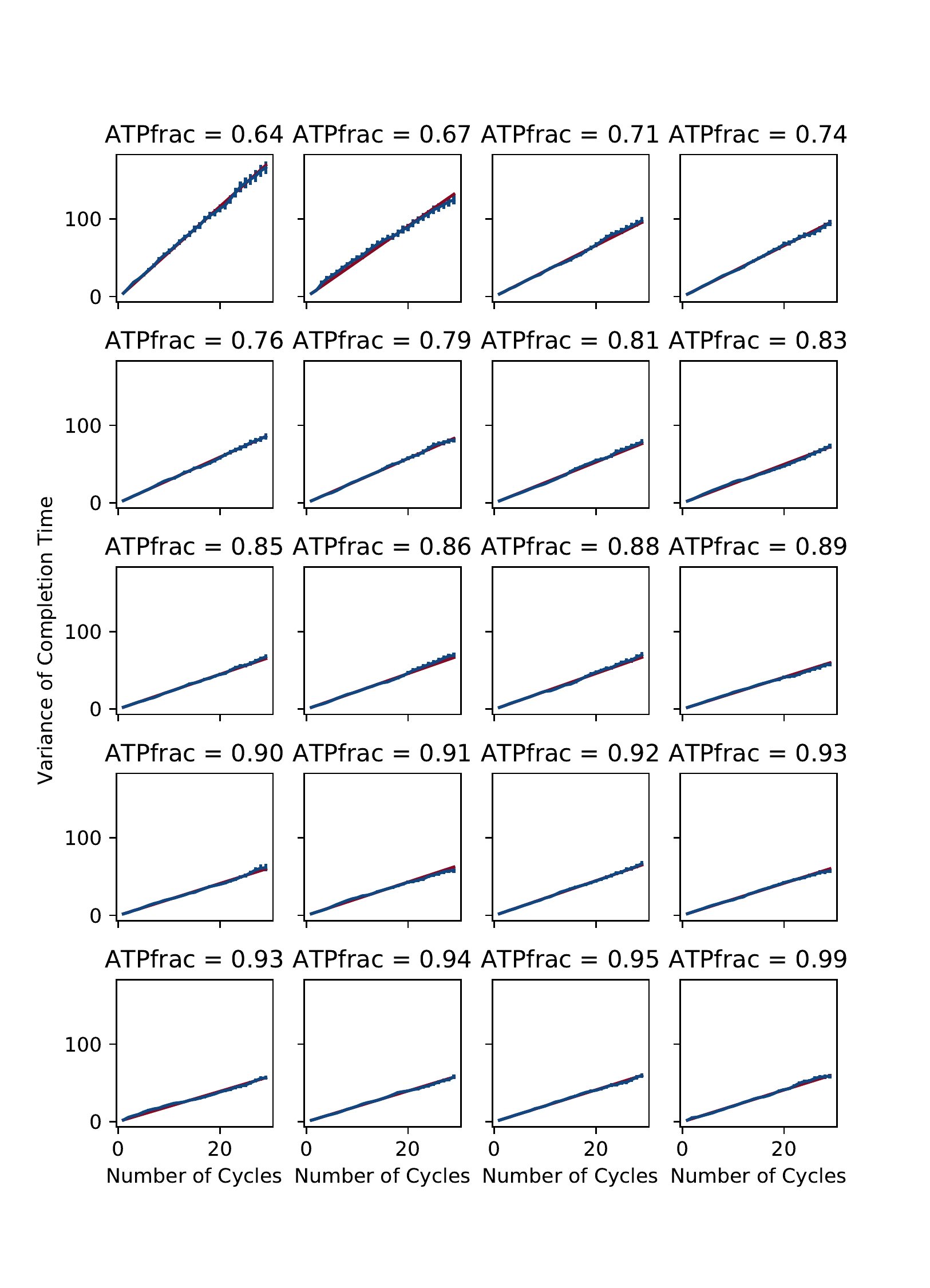}
	\caption{ {\bf Estimating $\mathcal{N}$ from KaiC simulations}. Each panel shows the estimated variance and bootstrapped 64\% confidence intervals in the $n$-cycle first passage time $\tau_n$. Straight black lines are linear fits, whose slopes provide the values of $D$ used in the computation of $\mathcal{N}$ for the main text figures.}
	\label{fig:fits}
\end{figure*}

After projecting each trajectory onto the first two principal components and computing the $n$-cycle first passage times as described above, we obtained the variance in $\tau_n$ as a function of $n$ for each ensemble. We computed bootstrapped 64\% confidence intervals for the estimate of the variance using the Python module ``bootstrapped,'' available at \url{https://github.com/facebookincubator/bootstrapped}. This data is plotted for all the KaiC ensembles in figure \ref{fig:fits}. The data is well fit by a straight line even for low $n$ in each of the plots. We obtained the slope $D$ of these lines using a weighted least-squares fit, also shown in the figure, with the weights provided by the inverse of the bootstrapped confidence intervals. 

We used these confidence intervals to obtain the bootstrap estimate for the uncertainty in $D$. The size of the confidence interval was proportional to $n$, as expected from a simple multiplicative noise model where the slope $D$ is a random variable. The constant of proportionality yields an estimate for the standard deviation of the distribution from which $D$ is sampled. We obtained this value for each data point with another least-squares linear fit, and used it to set the size of the error bars in figures \ref{fig:Neff} and \ref{fig:toy}.

%\section{Scaling with system size}
%
%\section{Renewal process approximation}
%
%In a renewal process, $\mathcal{N}$ could be computed from a single long trajectory, without explicitly generating an ensemble. This is because every cycle starts from the same state of the Markov process, and has no memory of former cycles. Thus ${\rm var}(\tau_n) = n {\rm var}(\tau_1)$, and each cycle in the trajectory generates an independent sample of $\tau_1$. 
%
%The KaiC model and our toy model are not renewal processes, because each cycle generically starts at a different point on the hyperplane that defines the winding number. To see whether this makes a difference to the computation of $\mathcal{N}$, we compared the data from the ensembles with data generated from $\tau_1$ measurements along single trajectories. Figure \ref{fig:renewal} shows that these two methods produce indistinguishable results for these models in the parameter regimes tested here.
%
%\begin{figure}
%%	\includegraphics[width=8cm]{figure7.pdf}
%	\label{fig:renewal}
%	\caption{ {\bf Validity of renewal process approximation} }
%\end{figure}

\end{document}